# Tuning the multiferroic mechanisms of TbMnO$_3$ by epitaxial strain

Kenta Shimamoto,[1] Saumya Mukherjee,[2] Sebastian Manz,[3] Jonathan S. White,[2] Morgan Trassin,[3] Michel Kenzelmann,[4] Laurent Chapon,[5] Thomas Lippert,[1,6] Manfred Fiebig,[3] Christof W. Schneider,[1,*] and Christof Niedermayer[2,*]

[1] *Laboratory for Multiscale Materials Experiments, Paul Scherrer Institut, CH 5232 Villigen-PSI, Switzerland*

[2] *Laboratory for Neutron Scattering and Imaging, Paul Scherrer Institut, CH 5232 Villigen-PSI, Switzerland*

[3] *Department of Materials, ETH Zurich, CH 8093 Zurich, Switzerland*

[4] *Laboratory for Scientific Development and Novel Materials, Paul Scherrer Institut, CH 5232 Villigen-PSI, Switzerland*

[5] *Institut Laue Langevin, BP 156X, 38042 Grenoble, France*

[6] *Laboratory of Inorganic Chemistry, Department of Chemistry and Applied Biosciences, ETH Zürich, CH 8093 Zurich, Switzerland*

[*]Correspondence and requests for materials should be addressed to C.N. and C.W.S. (email: christof.niedermayer@psi.ch, christof.schneider@psi.ch).

**Abstract**

A current challenge in the field of magnetoelectric multiferroics is to identify systems that allow a controlled tuning of states displaying distinct magnetoelectric responses. Here we show that the multiferroic ground state of the archetypal multiferroic $TbMnO_3$ is dramatically modified by epitaxial strain. Neutron diffraction reveals that in highly strained films the magnetic order changes from the bulk-like incommensurate *bc*-cycloidal structure to commensurate magnetic order. Concomitant with the modification of the magnetic ground state, optical second-harmonic generation (SHG) and electric measurements show an enormous increase of the ferroelectric polarization, and a change in its direction from along the *c*- to the *a*-axis. Our results suggest that the drastic change of multiferroic properties results from a switch of the spin-current magnetoelectric coupling in bulk $TbMnO_3$ to symmetric magnetostriction in epitaxially-strained $TbMnO_3$. These findings experimentally demonstrate that epitaxial strain can be used to control single-phase spin-driven multiferroic states.

Materials with both electric and magnetic order, called multiferroics[1], offer opportunities to magnetically control their electric properties and vice versa[2,3]. For device applications the magnitude of the mutual coupling between these orders is of crucial importance. One of the most widely studied materials with multiple ferroic properties is $BiFeO_3$. It exhibits multiferroicity at room temperature and the structural phase[4] and magnetic properties[5] are found to be modified by epitaxial strain when grown as thin films[6]. In such a system, despite the independent origin for their magnetism and ferroelectricity, a reorientation of the magnetic order can be achieved using electric fields in bulk[7,8] but also in thin films[9]. One can also expect a strong magnetoelectric coupling intrinsic to a compound itself in so-called spin-driven ferroelectrics and act on the electric and magnetic order with an external magnetic or electric field, respectively[10-12]. In the last 15 years, a number of spin-driven multiferroics have been discovered. However, it has remained a challenge to tune their physical properties. Recently it was shown that hydrostatic pressure in bulk $TbMnO_3$ can switch its multiferroic ground state by completely changing the magnetic phase, leading to a large enhancement of the ferroelectric (FE) polarization ($P$)[13,14]. For thin-film rare-earth manganites, it was shown that strain may alter the magnetic properties[15,16] leading to e.g. a complex coexistence of magnetic order parameters[15].

Here we demonstrate that the multiferroic ground state of a $TbMnO_3$ thin film can be tuned by epitaxial strain to adopt very different multiferroic phases. Our conclusions are supported by an extensive characterization of the magnetic and electric properties. Depending on the strain state, our $TbMnO_3$ films can adopt either a spin-spiral-induced ferroelectric ground state observed in the bulk[17,18], or a clean E-type magnetic ground state with large $P$. The stabilization of single-phase multiferroic phases in films, and their control using epitaxial strain, is an important milestone towards the development of device applications based on multiferroic films.

**Results**

TbMnO$_3$ films were prepared on (010) and (100) oriented YAlO$_3$ substrates by pulsed laser deposition using a KrF excimer laser. The lattice mismatch between TbMnO$_3$ and a (010) YAlO$_3$ substrate was ~ 2.1 % along the *a*- and ~ 0.4 % along the *c*-axis[19,20]. A (100) YAlO$_3$ substrate was used to prepare a relaxed reference sample expected to display bulk-like properties. Out-of-plane x-ray *θ*-2*θ* diffraction patterns (Fig. 1a) indicate the TbMnO$_3$ films grown on both (010) and (100) YAlO$_3$ to be single-phase without any twinning. (010) oriented TbMnO$_3$ films (14 and 44 nm) were obtained on (010) YAlO$_3$ substrates. A (100) oriented TbMnO$_3$ film was grown on a (100) YAlO$_3$ substrate different to previous reports[21,22]. In order to investigate lattice parameters of those films, reciprocal lattice maps were taken using a four-circle x-ray diffractometer (Fig. 1b-e, Table 1). The lattice of the (010) oriented 44 nm film is clamped to the substrate as demonstrated in the (130) and (041) reflections, exhibiting the same in-plane components of the reciprocal lattice points ($Q_a$ and $Q_c$). The peak locations clearly deviate from those estimated for bulk, verifying that the film is largely strained. The out-of-plane lattice parameter of the film is expanded by 1.7 % as a consequence of epitaxial strain. The 14 nm film showed the same crystallographic properties. Contrary to the (010) oriented film, the (100) oriented 80 nm film exhibited a relaxed structure, having the (402) and (310) reflections close to those of bulk and a fraction of *c*-axis strained layer. The (100) oriented film also displayed a large mosaicity as indicated by the broadened (402) and (310) peaks, while the (010) oriented film has very good crystallinity shown by the sharp (130) and (041) reflections with Laue oscillations. Hereafter, we refer to the (010) oriented films as "the strained films" and to the (100) oriented film as "the relaxed film." Using thick (010) oriented films is not the best choice to investigate properties of relaxed TbMnO$_3$ films by neutron diffraction and SHG since these thick films typically consist of multiple layers (strained, partially relaxed, and fully relaxed)[23] with modified physical properties for each layer. Those measurements probe signals from the entire film thickness and data cannot be analysed unambiguously.

The magnetic order in the films was investigated by neutron diffraction using the triple-axis spectrometer RITA-II at PSI, SINQ (Switzerland) and the single crystal four-circle diffractometer D10 at ILL (France). The films were aligned in the (0 $k$ $l$) scattering plane in order to access the strong magnetic reflection at (0 $q_k$ 1)[24]. Representative scans along (0 $k$ 1) at selected temperatures are displayed in Fig. 2. From these data we find the strained film to show a commensurate phase with $q_k = 0.50$ (r.l.u.) (Fig. 2a) below ~ 31 K which is in a sharp contrast to the relaxed film which exhibits bulk-like magnetic properties with an incommensurate (IC) magnetic wave vector with $q_k \sim 0.29$ (r.l.u.) from Mn spins (Fig. 2b)[24,25]. We do not observe any other peak between $0.2 \leq q_k \leq 0.55$ from the strained film at 15 K demonstrating that the magnetic order of the Mn spins is completely modified by epitaxial strain. The narrow peak width of the magnetic reflections of the strained film is close to the instrumental resolution (Fig. 2a inset) and reveals an out-of-plane magnetic correlation length of the order of 40 nm which corresponds to the film thickness. This is to the best of our knowledge the first report of an orthorhombic rare-earth manganite film showing a commensurate magnetic diffraction peak which directly implies E-type antiferromagnetism (AFM). The existence of E-type AFM claimed in previous reports is indirectly deduced from a large $P$ along the $a$-axis (∥$a$) and/or structural diffraction measurements[15,26].

In order to probe the FE state in the strained film, we performed SHG experiments. SHG is sensitive to inversion-symmetry-breaking FE order and a well-established tool to investigate these systems non-invasively[12,22,27]. First, from the temperature-dependence of the SHG response we observe a significantly increased FE transition temperature ($T_{FE}$) at ~ 41 K (Fig. 3a). This is about 15 K higher than $T_{FE}$ in the bulk-like relaxed films (see inset Fig. 3a and ref. [22]). Second, to verify that the $P$-direction flipped from ∥$c$ to ∥$a$ in the strained film, we performed light polarization dependent SHG measurements as shown in Fig. 3b. Here, we kept the incoming light polarization parallel to [100] while the outgoing polarization dependence was mapped out. The symmetry-based analysis confirms that the *P* points along the *a*-axis.

Furthermore, we observed a remarkable increase in the SHG intensity in the strained film. It is at least two orders of magnitude larger than in bulk[12] and significantly larger than in bulk-like films (Fig. 3a and ref. [22]). This enormous gain might be attributed to the corresponding strain induced *P* enhancement since the SHG intensity is proportional to $P^2$. Our SHG measurements therefore reveal that (i) the ordering temperature $T_{FE}$ as well as the SHG yield are substantially increased and (ii) the polarization direction flipped from the *c*-axis to the *a*-axis.

In Fig. 4 the temperature dependencies of the magnetic and electric properties of the strained films are summarized. The neutron scattering intensity of the peak measured at $q_k = 0.5$ is plotted in Fig. 4a and extrapolates to zero at approximately 41 K which is very close to the $T_{FE}$. The temperature dependence of the peak position of the magnetic order vector is shown in Fig. 4b. It is constant below $T_{lock} \sim 31$ K with a concurrent locking of the magnetic order into E-type AFM ($q_k = 0.5$) where Tb spins, too, are expected to exhibit an ordered state[14,24,25]. Capacitance measurements show a divergent behaviour at around 41 K which corresponds to the $T_{FE}$ (Fig. 4c). Thus, the strained film exhibits the following multiferroic phases: E-type AFM with ferroelectricity below $T_{lock}$ and IC AFM with ferroelectricity between $T_{lock}$ and $T_{FE}$. The magnitude of $P||a$ at 15 K is ~ 2 μC cm$^{-2}$ which is more than twenty-five times larger than bulk[17] and corresponds to the values estimated for *o*-HoMnO$_3$ with E-type AFM by the point charge model or by a specific DFT approach[28]. The stated value for *P* is larger than the reported bulk TbMnO$_3$ value of ~ 1 μC cm$^{-2}$ at 5 K and 5.2 GPa, and comparable to ~ 1.8 μC cm$^{-2}$ at $H = 8$ T[13], both values being considered to be among the highest ever reported for spin-driven FE.

**Discussion**

Our results show that highly strained high-quality TbMnO$_3$ films grown coherently on (010) oriented YAlO$_3$ substrates exhibit commensurate AFM with $P\|a$ as a ground state, while bulk shows an IC spin-spiral order with $P\|c$[17,18]. Meanwhile, $T_{FE}$ and the magnitude of $P$ are enhanced from 28 K to ~41 K and ~ 0.06 μC cm$^{-2}$ to ~ 2 μC cm$^{-2}$, respectively[29]. The observed change of the ground state in TbMnO$_3$ induced by two-dimensional growth-induced stress is by chance similar to the application of three-dimensional chemical pressure (i.e. substitution of smaller RE ions)[30] or hydrostatic pressure[13,14]. The modified ground state can be attributed to a strain-tuned dominant magnetoelectric coupling mechanism, from antisymmetric magnetostriction (inverse Dzyaloshinskii-Moriya interaction)[31-33] to symmetric magnetostriction[33-35]. Our results thus clearly demonstrate a strain-induced tuning of a dominant magnetoelectric coupling mechanism in multiferroic materials, presenting a unique way to control its physical properties.

At the microscopic level, the role of epitaxial strain can be interpreted as follows. The lattice of the strained film is compressed along the *a*-axis by 2.1 % and expanded along the *b*-axis by 1.7 %. Hence, it is expected that the distance between the two oxygen atoms which mediate the next-nearest-neighbour exchange interaction along the *b*-axis ($J_b$) (O(2) and O(3) in Fig. 2c) becomes smaller than in the bulk (≈ the relaxed film). This leads to a larger orbital overlap between those oxygen ions and, consequently, $J_b$ increases. The increase of $J_b$ seems to be key to trigger the symmetric magnetostriction[18]. The large $P$ enhancement by epitaxial strain implies a significant increase of a Peierls-type spin-phonon coupling and/or a reduction of elastic energy for the shift of atoms, which also contribute to stabilize E-type AFM by increasing the magnitude of biquadratic interaction[33,35-37]. Since there is no direct access to the experimental verification of the location of oxygen in the film so far to obtain values for those interaction parameters, *ab-initio* calculations are at present the only way to verify the validity of the abovementioned hypothesis.

Unlike in other reported orthorhombic rare-earth manganites that induce FE order simultaneously with E-type AFM ($T_{lock} = T_{FE}$)[38,39], our strained TbMnO$_3$ film exhibits ferroelectricity while its magnetic order is still IC ($T_{lock} < T_{FE}$). One possible candidate for such a phase is a mixture of stable E-type and meta-stable IC AFM as suggested from Monte Carlo simulations[33,35]. According to the calculated phase diagram, the IC AFM may disappear at low temperatures depending on the magnitude of $J_b$, which also fits to our observation of E-type AFM below $T_{lock}$. Here we note that neither the SHG signal nor the capacitance shows an anomaly in their temperature dependencies at around $T_{lock}$ (Figs. 3a and 4c), i.e. the temperature variation of the magnetic order at around $T_{lock}$ seems not to affect the FE properties. This feature contradicts the FE properties as calculated by the Monte Carlo simulations where an abrupt increase of $P$ is expected when an IC component disappears[35]. Another potential explanation is that the spins are ordered in an *ab*-cycloidal structure with a very long periodicity. In such cases, a symmetric magnetostriction mechanism can still be dominant. Further studies are required to understand the magnetic structure of the phase with an IC magnetic diffraction peak between $T_{FE}$ and $T_{lock}$.

In summary, we demonstrated the modulation of the multiferroic mechanism in TbMnO$_3$ using epitaxial strain. Films coherently grown on (010) oriented YAlO$_3$ substrates are strongly strained and exhibit a commensurate magnetic diffraction peak with a strongly enhanced ferroelectric polarization oriented along the *a*-axis. In contrast a relaxed TbMnO$_3$ film prepared on a (100) oriented YAlO$_3$ substrate shows bulk-like structural and multiferroic properties. The ground state of the strained film represents the emergence of a dominant symmetric magnetostriction which is absent in bulk and the relaxed film. The microscopic origin can be attributed to the strain-driven enhancement of the next-nearest-neighbour exchange interaction between Mn ions along the *b*-axis.

**Methods**

**Sample preparation and structural characterization**

Epitaxial films of TbMnO$_3$ are grown on (010) oriented YAlO$_3$ single crystalline substrates by pulsed laser deposition using a KrF excimer laser ($\lambda$ = 248 nm, 2 Hz). The laser beam is focused onto a sintered ceramic target with a spot size of ~ 1.2 × 1.7 mm. The laser fluence was adjusted to 2.0 J cm$^{-2}$. The substrate is located on-axis to the plasma plume with a distance of 4.1 cm from the target. Deposition was performed in an N$_2$O background at 0.7 mbar with the substrate heated to 690°C by a lamp heater[23]. The reference TbMnO$_3$ film on a (100) oriented YAlO$_3$ substrate, too, was prepared by pulsed laser deposition with a different heater and conditions. A Si resistive heater maintained the temperature at 760°C during the growth with the target-substrate distance of 3.7 cm and N$_2$O background at 0.3 mbar. Right after the deposition the sample was cooled in the same gas environment as the film growth. Reciprocal space maps of films are taken by using a Seifert four-circle x-ray diffractometer with Cu x-ray source equipped with monochromator.

**Neutron diffraction measurements**

The neutron diffraction measurements carried out at the neutron triple-axis spectrometer RITA-II, SINQ, PSI, utilized an incident wavelength of $\lambda$ = 4.21 Å obtained from the (002) Bragg reflection of a vertically focusing pyrolytic graphite (PG) monochromator. A PG filter between the monochromator and the sample, and a cooled Be filter between the sample and the analyser were installed to suppress the higher order contamination. The sample was mounted in the (0$k$0)-(00$l$) scattering plane. RITA-II is equipped with a nine-bladed PG analyser, which provides a high $q$-resolution and improves the signal to noise ratio. For the diffraction experiment, the central blade is used. To ensure a collimated incident beam an 80' external collimator was installed between the monochromator and the PG filter. In experiments conducted at D10, ILL, an incident wavelength of $\lambda = 2.364$ Å was used without any collimator before or after the sample. Vertical focusing PG crystals were used as an analyser. The four-circle diffractometer both provides an access to a broad range of *hkl* scattering planes from the sample, and is equipped with an advanced He-4 cryostat for providing cryogenic sample temperatures.

**Second-harmonic generation measurements**

SHG is a nonlinear optical process denoting the emission of light at frequency $2\omega$ from a crystal irradiated with light at frequency $\omega$. This is expressed by the equation $P_i(2\omega) = \varepsilon_0 \Sigma_{j,k} \chi^{(2)}_{ijk} E_j(\omega) E_k(\omega)$, where $E_{j,k}(\omega)$ and $P_i(2\omega)$ are the electric-field components of the incident light and of the nonlinear polarization, respectively, with the latter acting as the source of the SHG wave. The nonlinear susceptibility $\chi^{(2)}_{ijk}$ characterizes the ferroelectric state.

Multiferroic TbMnO$_3$ possesses the point group symmetry *mm2* (*2*-axis || *P*). For a spontaneous polarization along the *c*-axis, the relevant SHG tensor components then yield $\chi^{(2)}_{ccc}$, $\chi^{(2)}_{caa}$ and $\chi^{(2)}_{aca}$. In the strained phase the polarization reorients along the *a*-axis with the dominant tensor component $\chi^{(2)}_{aaa}$.

For probing the TbMnO$_3$ films, we used light pulses emitted at 1 kHz from an amplified Ti:sapphire system with an optical parametric amplifier. The light pulses had a photon energy of 1.0 eV, a pulse length of 120 fs and a pulse energy between 2 - 20 µJ. Detailed technical aspects of SHG in ferroic systems and especially in TbMnO$_3$ are described in refs. [12,27].

**Electrical characterization**

In order to evaluate in-plane electric properties of films, Au (56 nm)/ Ti (4 nm) interdigitated electrodes were patterned on the film surface by photolithography and lift-off procedures. The finger width and gap are 5 µm and the line length is 1.25 mm. Measurements were performed at continuous helium flow atmosphere and temperature was controlled by a LakeShore Model 325 temperature controller. Capacitance measurements were performed using an Agilent E4980A LCR meter at zero DC field with an AC voltage of 100 mV. The frequency is varied from 100 to 2 MHz and the data taken at 15 kHz are shown in Fig. 4c. A ferroelectric hysteresis curve was probed through the Positive-Up Negative-Down (double-wave) method[40], using National Instruments compact DAQ analog input (NI 9229)/ output (NI 9263) modules and a home-made Sawyer-Tower circuit. The frequency of the input sinusoidal waves was set to 1 kHz. The polarization ($P$) was calculated as $P = Q(tL)^{-1}$ [41,42], where $Q$ is the measured charge, $t$ is the film thickness, and $L$ is the total length of the finger pairs.

**Acknowledgements**

This work is based on experiments performed at the Swiss spallation neutron source SINQ, Paul Scherrer Institute, Villigen, Switzerland. Financial support and CROSS funding to K.S. from PSI are acknowledged. S.Mukherjee acknowledges financial support from the Swiss National Science Foundation (SNF, project number 200021_147049) and J.S.W. from MaNEP and SNF (No. 200021_138018). M.T. and M.F. acknowledge funding through the SNSF R'Equip Program (Grant No. 206021-144988) and the EU European Research Council (Advanced Grant 694955 - INSEETO). We would like to thank M. Bator for providing the (100) oriented $TbMnO_3$ film, D. Marty (PSI, Laboratory for Micro- and Nanotechnology) for support with optical lithography, U. Greuter for the implementation of a Sawyer-Tower circuit for ferroelectric hysteresis measurements, and A. Scaramucci for a fruitful discussion. The drawings of crystal structures are produced by VESTA program[43], which is acknowledged.

**Author Contributions**

K.S. and S.Mukherjee contributed equally to this work. The strained films were prepared by K.S. along with structural and electric characterizations of all the samples. Neutron diffraction data were collected by S.Mukherjee, J.S.W., L.C., and C.N., and analysed by S.Mukherjee, J.S.W., and C.N. SHG measurements were conducted by S.Manz under the supervision of M.T. and M.F. The study was planned by K.S. and S.Mukherjee and supervised by M.K., T.L., C.W.S., and C.N. The manuscript was prepared by K.S., S.Mukherjee, and S.Manz with input from all co-authors.

**Competing Financial Interests**

The authors declare that they have no competing financial interests.

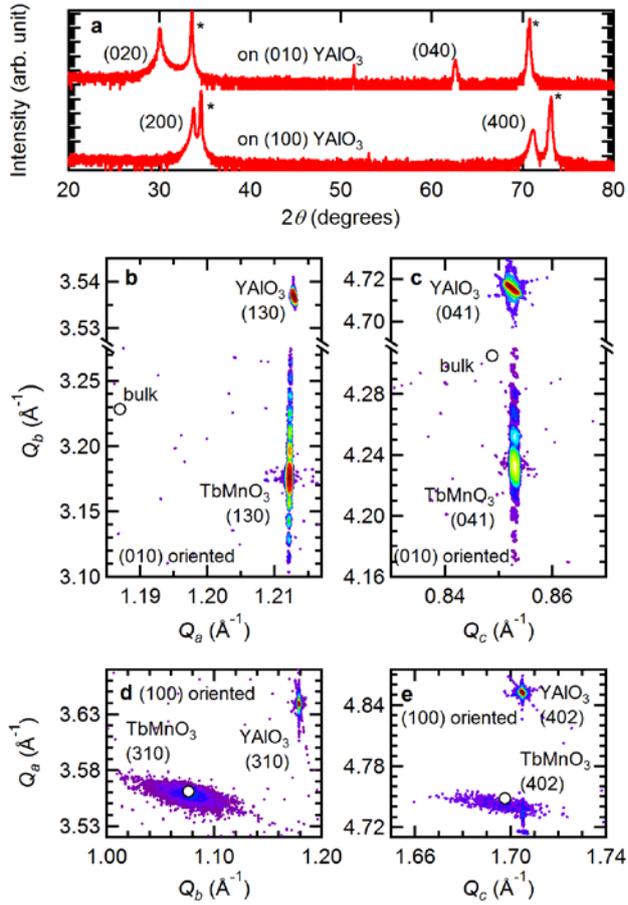

Figure 1. **Structural properties of TbMnO₃ films grown on a (010) and a (100) oriented YAlO₃ substrate.** (a) $\theta$-$2\theta$ scans of TbMnO$_3$ films on a (010) oriented (top) and a (100) oriented YAlO$_3$ (bottom) substrate. Each peak marked by an asterisk is from the YAlO$_3$ substrate. Reciprocal lattice maps of (b) the (130) and (c) the (041) reflection of a 44 nm TbMnO$_3$ film on a (010) oriented YAlO$_3$ substrate. Those of the (310) and the (402) reflection of a 80 nm TbMnO$_3$ film on a (100) oriented YAlO$_3$ substrate are shown in (d) and (e) respectively.

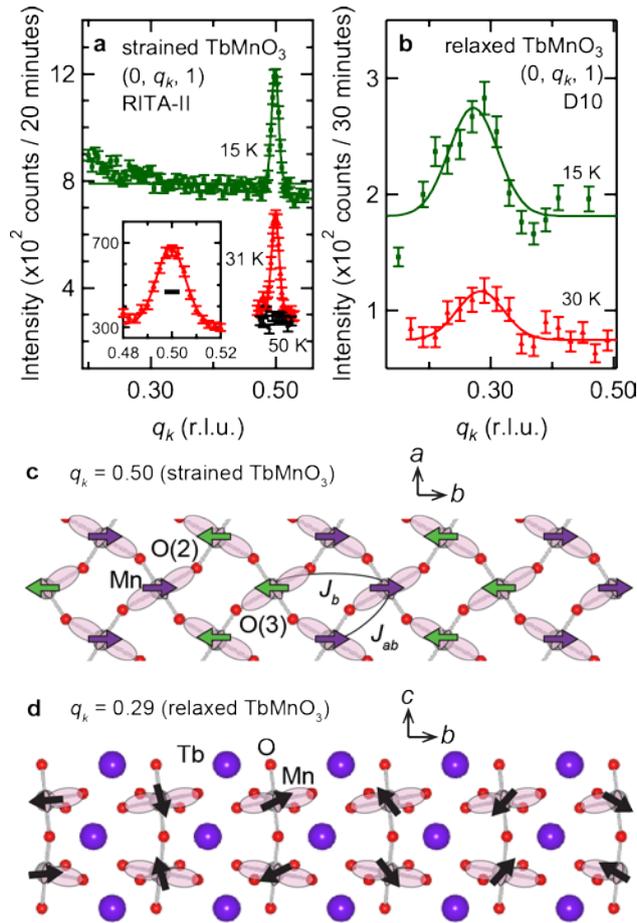

Figure 2. **Magnetic diffraction measurements of the TbMnO₃ films.** The $(0, q_k, 1)$ magnetic Bragg reflections measured at 15 and ca. 30 K of (a) the strained TbMnO$_3$ film and (b) the relaxed TbMnO$_3$ film. The $(0\ q_k\ 1)$ reflection at 31 K of the strained TbMnO$_3$ film is magnified in the inset of (a) and the black line marks the instrumental resolution. Data have been shifted for clarity. The black line marker gives the instrumental resolution. Schematic images of Mn spin order (c) in the *ab*-plane for E-type AFM and (d) in the *bc*-plane for *bc*-cycloid[18,44].

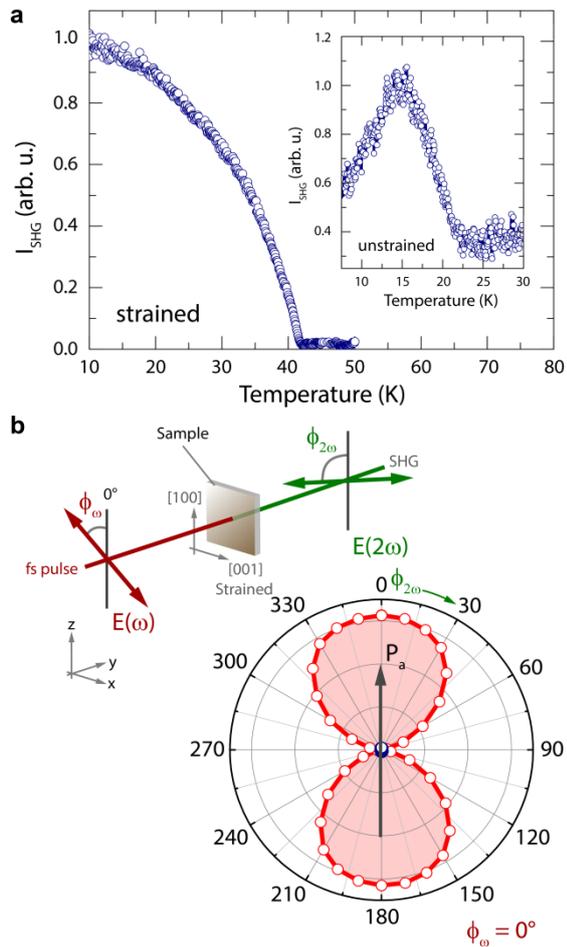

Figure 3. **Structural polar responses of the TbMnO$_3$ films.** (a) Temperature dependence of the SHG response of the strained TbMnO$_3$ film. The inset shows the corresponding data for the unstrained film. Both data sets were normalized to the maximum of the SHG intensity of the strained film. (b) SHG experimental geometry: The incoming light polarization of the fundamental beam at ω can be set and the outgoing frequency-doubled response at 2ω can be read out, respectively. The data points correspond to a measurement with the incoming polarization fixed along the (100) direction. The data matches the expected symmetry for a polarization pointing along the *a*-axis.

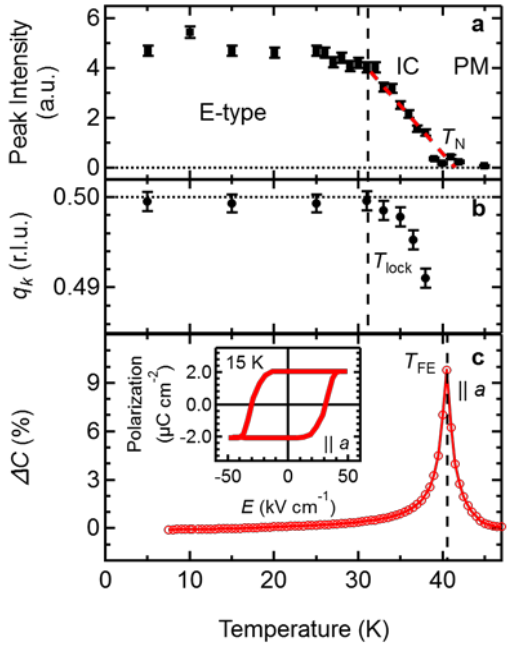

Figure 4. **Temperature dependent multiferroic properties of the strained TbMnO3 films.** Temperature dependent magnetic and electric properties of the strained TbMnO3 film. (a) Peak intensity at (0 0.5 1) magnetic reflection. A red dashed line is a guide to eye. (b) Peak position of the (0 $q_k$ 1) magnetic reflection. (c) Normalized capacitance ($\Delta C = (C(T) - C(50\,K))/C(50\,K)$) measured along the *a*-axis. The inset shows a ferroelectric hysteresis curve at 15 K. Panel (c) shows data obtained on the 14 nm (010) TbMnO3 film.

TABLE I: Lattice parameters and strain (compressive, +; tensile, -) of TbMnO$_3$ films derived from Fig. 1.

|  | YAlO$_3$[19] | TbMnO$_3$ bulk[20] | The (010) film (strained) | The (100) film (relaxed) |
|---|---|---|---|---|
| a (Å) | 5.180 | 5.293 | 5.182 | 5.30 |
| (strain) |  |  | + 2.1 % | - 0.1 % |
| b (Å) | 5.329 | 5.838 | 5.936 | 5.82 |
| (strain) |  |  | - 1.7 % | + 0.3 % |
| c (Å) | 7.371 | 7.403 | 7.371 | 7.41 |
| (strain) |  |  | + 0.4 % | - 0.1 % |